# Modeling tumorspheres reveals cancer stem cell niche building and plasticity

*L. Benítez, L. Barberis, C. A. Condat*

Instituto de Física Enrique Gaviola, CONICET, and
Facultad de Matemática, Astronomía, Física y Computación,
Universidad Nacional de Córdoba,
Ciudad Universitaria, X5000HUA Córdoba, Argentina.

**Abstract**

Cancer stem cells have been shown to be critical to the development of a variety of solid cancers. The precise interplay mechanisms between cancer stem cells and the rest of a tissue are still not elucidated. To shed light on the interactions between stem and non-stem cancer cell populations we develop a two-population mathematical model, which is suitable to describe tumorsphere growth. Both interspecific and intraspecific interactions, mediated by the microenvironment, are included. We show that there is a tipping point, characterized by a transcritical bifurcation, where a purely non-stem cell attractor is replaced by a new attractor that contains both stem and differentiated cancer cells. The model is then applied to describe the outcome of a recent experiment. This description reveals that, while the intraspecific interactions are inhibitory, the interspecific interactions stimulate growth. This can be understood in terms of stem cells needing differentiated cells to reinforce their niches, and phenotypic plasticity favoring the de-differentiation of differentiated cells into cancer stem cells. We posit that this is a consequence of the deregulation of the quorum sensing that maintains homeostasis in healthy tissues.

**Keywords**: tumorsphere, cancer stem cell, plasticity, spheroid, mathematical model

## 1. Introduction

The cancer stem cell hypothesis has been continuously gaining ground with the identification of cancer stem cells (CSCs) in a growing number of tumor types [Lapi94, Hajj03, Singh04, Li07, Obrien07, Eramo08]. It states that cancer growth is driven by a subpopulation of stem cells and helps to explain tumor recurrence and metastasis [Batll17]. It has also led to new therapeutic paradigms, based on the concept that destroying or incapacitating CSCs would be an efficient method of cancer containment and control. Cancer stem cell targeted vaccines are also being developed [Lin17].

Stem cells are associated to niches, sets of cells and intercellular elements that provide the signals that define stem cell behavior and maintain stemness [Schof78, Fuchs04, Scad14]. Although the functionality of the cancer stem cell niche is well established, its structure and design have no precise definition. Its deregulation could be a hallmark of tumorigenesis. In normal tissues, signals provided by the niche lead to the maintenance of homeostatic equilibrium. Under equilibrium conditions, the proportion of stem cells is relatively small. This equilibrium is likely to be destroyed in cancerous tissues, which may exhibit a larger ratio of stem cells [Chen16]. It has been observed that in high-grade tumors from patients, an increased pool of CSCs may underlie their aggressive behavior [Visv12, Al-Hajj04].

The obvious importance of CSCs has led to the formulation of various mathematical models of CSC-fueled tumor growth. Discrete models for chronic leukemia [Ding08] and solid cancers [Lapor12] describe the early stages of disease progression, while statistical models estimate probabilities, e.g. to acquire drug resistance [Boma07, Toma10], and computational simulations [Gao13, Ende09, Wacl15] model macroscopic behavior starting from microscopic interaction rules. Diverse continuous population dynamic models have also been developed [Gang06, Liu13]. These focus variously on feedback loops [Rodri11], delayed interactions [Ghos17], stability analyses [Guo17] or bifurcation landscapes [Ferr12, Guis17]. These models offer insights into growth and differentiation rates, cell population fractions, lateral inhibition, and chemo- and radio-therapy effects, to cite a few examples.

Our approach here is to develop a continuous model that includes environment- mediated cell-cell interactions in a classic predator-prey formulation. To keep the model as simple as possible, we explicitly include only stem and differentiated cancer cells. It is then suitable to describe the growth of tumorspheres. A tumorsphere (reviewed in [Weis15]) is a simplified biological system that enables the study of cancer progression in the absence of healthy tissue and of the complex signaling system prevalent in the environment in which real tumors live. The term tumorsphere specifically refers to spheroids grown from single-cell suspensions out of permanent cell lines, tumor tissue, or blood cultivated in a serum-free medium with EGF and FGF-2 growth factors and characterized by low-attachment plastic, clonal density, and potential preliminary cell sorting. Thus tumorspheres are formed by clonal proliferation in low-adhesion conditions and with stem cell medium [Weis15]. We therefore expect that the application of the model to tumorsphere experimental data will yield information about some of the basic properties of an interacting system of stem and differentiated cancer cells. In their study of mammary tumorspheres, Gu *et al.* observed that, under certain conditions, stem cells "underwent differentiation induced by environmental stimuli" [Gu11]. It is therefore conceivable that the final status of a tumorsphere be composed of differentiated cells alone. Our model will help us to identify the conditions under which this may occur.

The microenvironment also regulates cell plasticity, the ability to reversibly transition between non-stem and stem cell states [Batll17]. This is an important feature that may seriously complicate anti-stem-cell therapy, since destroyed cancer stem cells may be replaced by dedifferentiating non-stem cancer cells. Moreover, the phenotypic diversity in a cancer cell population [Zhou14] and the diverse conditions under which experiments are performed [Viei13] may hinder the analysis of plasticity. Since tumorspheres may be created under controlled conditions and have only two subpopulations (stem and differentiated cancer cells) they can be used as simple biological models to identify the effects of plasticity. Indeed, the application of our model to an experiment suggests that even in the relatively simple environment of a tumorsphere, plasticity may play a crucial role in the system growth. In an interesting experiment, Gupta and coworkers isolated populations displaying stem cell, basal-, or luminal-like phenotypes from breast cancer cell lines. In vitro, all three subpopulations were able to generate cells of the other two phenotypes, such that the cultures converged over time toward the proportions of cell types observed in the original cancer line [Gup11]. The implication of this and other experiments is that stochastic events can preserve a dynamic equilibrium between stem and non-stem states [Sree18].

Since the model discussed in this paper does not include spatial variables, we will not be able to predict the system architecture, but we can still predict its growth dynamics and overall composition. Adhesion tends to keep dividing cells together and cancer cell groups in vitro form globular systems. We will thus assume that, if enough cells are created, the system we are studying becomes a tumorsphere.

In what follows we describe our two-population ecological modeling of the tumor and analyze its mathematical properties. We find that the final tumorsphere may be composed either of a mixture of stem and differentiated cancer cells or of differentiated cancer cells alone. We also find that the tumorsphere size increases with the fraction of cancer stem cells in the final state. Next we use the model to describe the results of an experiment of tumorsphere growth [Chen16] and conclude with a discussion of the implications of our findings about the interaction parameters.

## 2. Mathematical model

Cancer stem cells may divide symmetrically, generating two CSCs, asymmetrically, generating one CSC and one differentiated cancer cell (actually one progenitor cell is generated instead, but since this cell is committed to differentiate, we will omit it in our modeling), or they may generate two differentiated cells [Batll17].

We model the growth of a tumorsphere considering two cell populations: Cancer stem cells ($S$) and differentiated cancer cells ($D$). By including in the last class all cells with any degree of differentiation we can isolate the role played by stem cells. We will call the basal growth rates, i.e. the number of cell divisions per unit time, of differentiated and stem cells $r_D$ and $r_s$, respectively. We further assume that,

- The members of each subpopulation interact with each other (intraspecific interactions) and with the members of the other subpopulation (interspecific interactions). These interactions are quantified by a matrix of coefficients, $\alpha_{ij}$, whose diagonal elements describe intraspecific interactions and whose off-diagonal elements correspond to interspecific interactions.

- When a CSC undergoes mitosis there is a probability $p_s$ that two new S cells are generated and a probability $p_d$ that two D cells are generated. Because of normalization, the probability that there is an asymmetric division is $p_a = 1 - p_d - p_s$.

We can describe the evolution of the two interacting populations by generalizing the standard equations for two competing species (see, p. ej. [Brit03], p. 67), as follows,

$$\frac{dS}{dt} = r_S(p_s - p_d)S - r_S p_s S(\alpha_{SS}S + \alpha_{SD}D), \tag{1a}$$

and

$$\frac{dD}{dt} = \{r_D D + r_S S[1 - (p_s - p_d)]\}[1 - (\alpha_{DD}D + \alpha_{DS}S)]. \tag{1b}$$

Note that asymmetric divisions do not change the number of CSCs. The first term on the right-hand side of Eq. (1a) expresses the contribution of the intrinsic cell dynamics to the fate of a dividing CSC and is uncorrelated to the rest of the tissue. The first factor on the right-hand side of Eq. (1b) must be understood as the rate of creation of differentiated cells due to the division of $D$ cells, the asymmetric division of $C$ cells, and the differentiation of $C$ cells. The last parts of both equations indicate that the niche is not completely deregulated. Note that $\alpha_{ij} < 0$ describes a *cooperative* interaction that promotes growth. Conversely, positive values of $\alpha_{ij}$ describe growth-inhibiting competition. Since plasticity is regulated by the microenvironment, it cannot be described by the intrinsic growth rates, but is expressed through the interaction coefficients $\alpha_{ij}$.

Of course, our choice of Eqs. (1) is not unique. In the Appendix we develop a variant of this model, where the rate of cancer stem cell division is strongly influenced by the rest of the tissue, perhaps through a quorum sensing mechanism.

## 2.1. Non-dimensionalization

To better understand the underlying system dynamics, we non-dimensionalize Eqs. (1) using the new variables,

$$X = \alpha_{DD}\frac{r_S}{r_D}(1 - p_s + p_d)S$$

$$Y = \alpha_{DD}D$$

$$r = r_D t$$

$$P = \frac{r_S}{r_D}(p_s - p_d) \tag{2}$$

$$A = \frac{\alpha_{SS}}{\alpha_{DD}}\frac{p_s}{1 - p_s + p_d}$$

$$B = \frac{\alpha_{SD}}{\alpha_{DD}}\frac{r_S}{r_D}p_s$$

$$C = \frac{\alpha_{DS}}{\alpha_{DD}}\frac{r_D}{r_S}\frac{1}{(1 - p_s + p_d)}$$

We thus obtain the following set of dimensionless equations:

$$\frac{dX}{d\tau} = X(P - AX - BY) \qquad (3a)$$

$$\frac{dY}{d\tau} = (X + Y)(1 - CX - Y) \qquad (3b)$$

Since S ≥ 0 and D ≥ 0, their dimensionless counterparts X and Y must have the same sign.

## 2.2. Fixed points and bifurcation analysis

We are interested in CSC - driven growth, i.e., when there is net initial stem cell growth, which requires $P \geq 0$. We consider only this case in what follows. The relevant fixed points of model (3) correspond to three biological situations:

- The trivial point $P_0 = (0,0)$ which is always unstable. This means that, once a cell of any type is seeded, a colony will grow.

- A non-CSC fixed point $P_1 = (0,1)$, which is stable as long as P < B. The corresponding eigenvalues are $\lambda_{11} = -1$ and $\lambda_{12} = P - B$.

- A coexistence fixed point $P_2$, for which finite numbers of stem and differentiated cancer cells co-occur. This point is located at

$$P_2 = \frac{1}{A - BC}(P - B, A - PC) \qquad (4)$$

- There is a fourth fixed point,

$$P_3 = \left(\frac{P}{A - B}, -\frac{P}{A - B}\right), \qquad (5)$$

but this point is biologically irrelevant because one of the populations would be negative.

Since the eigenvalues corresponding to $P_2$ are too cumbersome to lend themselves to a simple stability analysis, we assume that this point becomes unstable if $P_1$ is stable. This assumption is supported by extensive numerical analyses. When $B = P$ the system undergoes a transcritical bifurcation as shown in Fig. 1. If $B < P$, the stable state is composed of a mixed population that contains $(X^*, Y^*)$cells, with,

$$X^* = \frac{(P - B)}{\Delta} \qquad (6a)$$

and

$$Y^* = \frac{(A - PC)}{\Delta}. \qquad (6b)$$

with $\Delta = A - BC$.

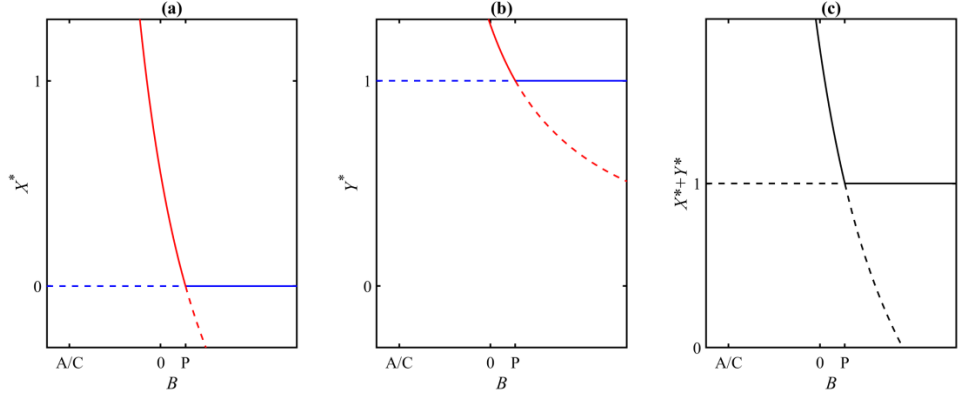

**Fig. 1**: Bifurcation diagram for **(a)** dimensionless CSC populations and **(b)** dimensionless DCC populations. Continuous lines correspond to stable branches, and dashed lines to unstable branches. Blue and red lines indicate the locations of the $P_1$ (pure DCC) and $P_2$ (coexistence) fixed points, respectively. The system undergoes a transcritical bifurcation at $B = P$ changing from a pure DCC spheroid to a mixed population coexistence. **(c)** Bifurcation diagram for the total cell population $X^* + Y^*$ in the steady state regime. Parameter values: $P = 1.1$, $C = -0.5$, and $A = 2$.

The parameters chosen for Fig. 1 correspond to $C < 0 < A$. The condition $C < 0$ means that one of $\alpha_{DD}$ and $\alpha_{DS}$ is positive and the other negative, while $A > 0$ means that both intraspecific interactions have the same sign. Figure 2 depicts the system evolution using the dimensionless variables $(X, Y)$. Growth starts from a cancer stem cell seed, $(X_0, Y_0) = (0.01, 0)$. When $P < B$, all trajectories converge to the differentiated cancer cell fixed point, while trajectories with $P > B$ converge to the coexistence fixed point. By increasing the ratio $P/B$, the coexistence fixed point moves towards higher values of both populations. If we decrease $P/B$, we observe that the stable coexistence point $P_2$ moves toward the saddle point $P_1$. When $P_2$ goes through $P_1$, stability changes: $P_2$ enters the second quadrant, becoming biologically irrelevant, and all trajectories die in $P_1$.

For all values of $P/B$ there is an initial stage for which both populations grow. At later times, the number of CSCs decreases monotonically, except for values of $B$ below a threshold $B^*$ ($B^* = 0.415$ for the case of Fig. 2). For these values of $B$ both populations increase monotonically, arriving at the fixed point specified by Eqs. (6).

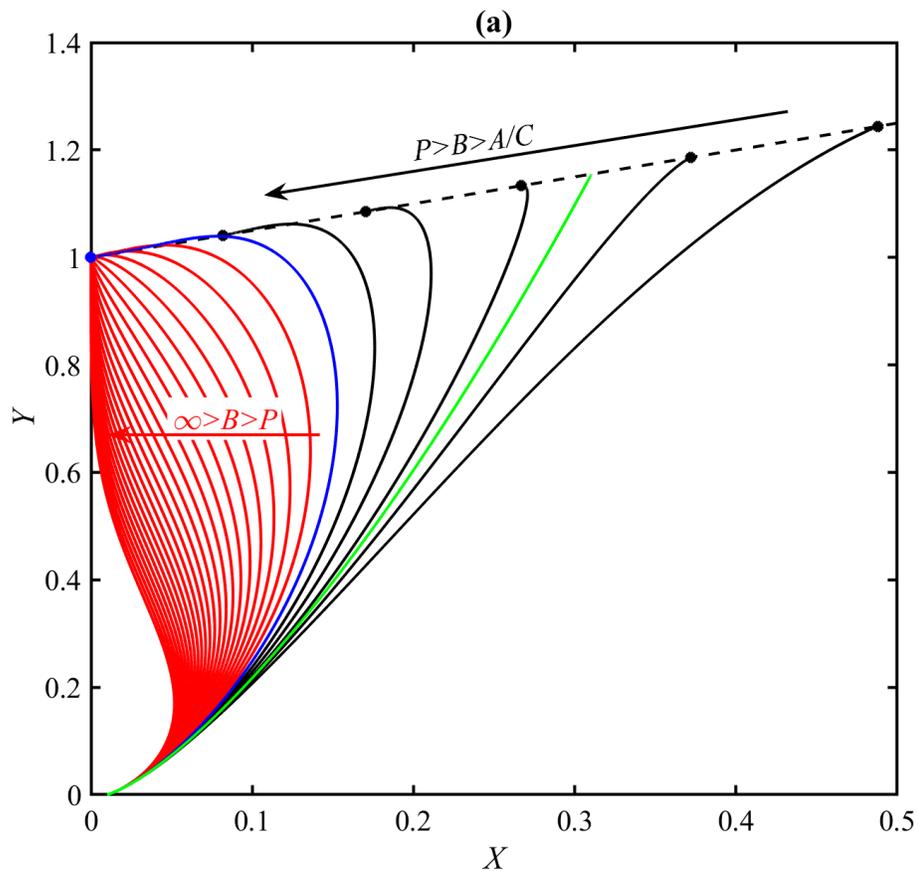

**Fig. 2a:** Trajectories in the $X - Y$ plane for an initially small stem cell population ($X_0 = 0.01, Y_0 = 0$). The blue trajectory corresponds to the bifurcation condition $B = P$. Black lines correspond to stable trajectories leading to coexistence. Red trajectories lead to the DCC fixed point. Arrows indicate the growth direction of parameter $B$.

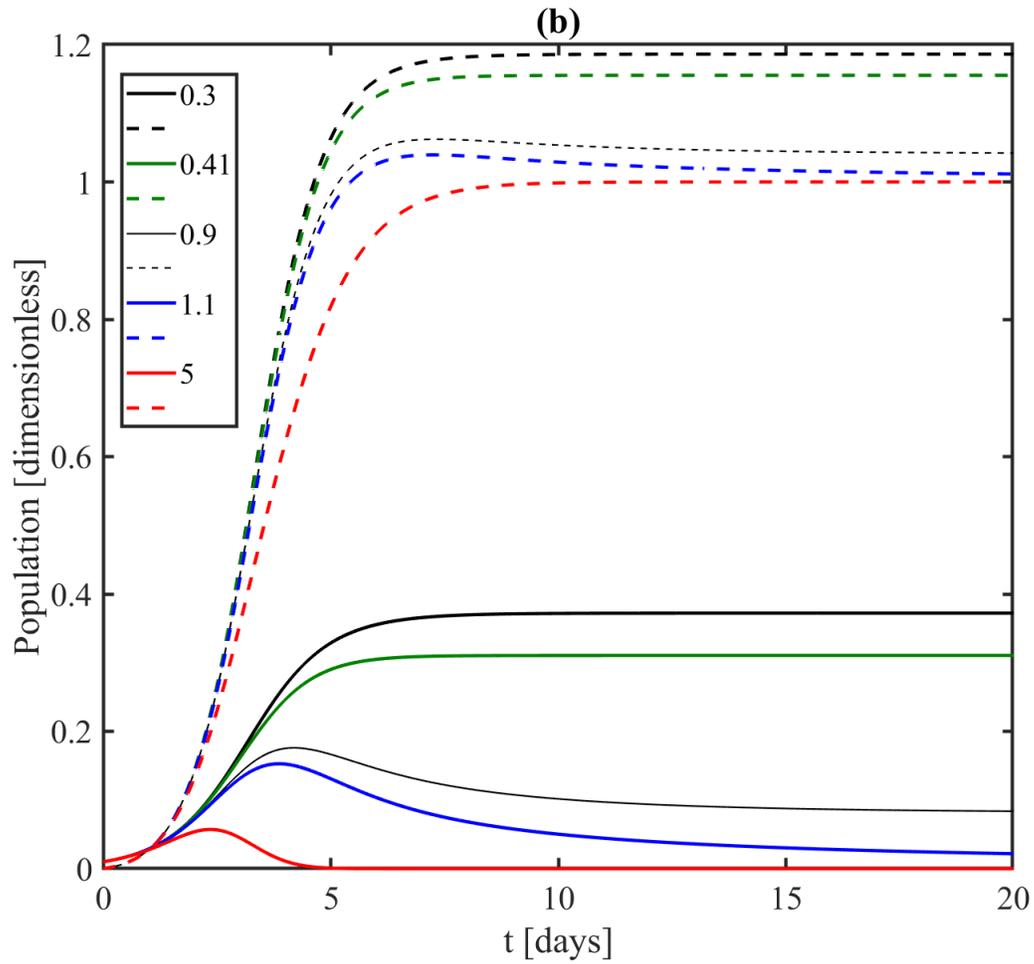

**Fig. 2b:** Time evolution of the cell populations for the values of $B$ indicated in the figure. Continuous (dashed) lines correspond to CSCs (DCCs). Red lines correspond to a final state of pure DCCs, green lines correspond to the threshold $B^*$, blue lines to the bifurcation, black and grey lines to coexistence. Parameter values: $P = 1.1$; $C = -0.5$, and $A = 2$.

## 2.3. Implications for the physical system.

Next, we examine the implications of our analysis in terms of the original functions $S(t)$ and $D(t)$. The stability condition ($P < B$) for the pure differentiated cancer cell attractor then reads,

$$\frac{\alpha_{SD}}{\alpha_{DD}} > 1 - \Pi, \qquad (7)$$

with $\Pi \equiv p_d/p_s \leq 1$. If the system parameters satisfy this condition, the resulting tumor is composed of $D_1 = 1/\alpha_{DD}$ differentiated cancer cells. The condition that the end system is finite demands that $\alpha_{DD} > 0$, i.e., that the intraspecific interaction between differentiated cancer cells be inhibitory. But, then, Eq. (7) requires that $\alpha_{SD} > 0$, i.e. that $D$ cells also inhibit $S$ cell growth. As expected, $D_1$ is a monotonically decreasing function of inhibition strength. Note also that for the special case $p_d = 0$ (no divisions generating two differentiated cells), Eq. (7) implies $\alpha_{SD} > \alpha_{DD}$: Differentiated cells must generate a strong inhibition of the cancer stem cell subpopulations. If $\alpha_{DD} \leq 0$ intraspecific stimulation causes the system to grow without limit. If $p_d/p_s = 0$, $\alpha_{SD} > \alpha_{DD}$: differentiated cells must exert a strong inhibition on the cancer stem cells. As $p_d/p_s$ increases (the relative probability of symmetric stem cell division decreases), a smaller value of the inhibition $\alpha_{SD}$ is possible. Finally, if $p_d/p_s > 1$, a range of negative (stimulatory) values of $\alpha_{SD}$ is compatible with limited growth.

Since $\lambda_1 = -1$, $P_1$ becomes a saddle point when $\lambda_2 > 0$. The coexistence fixed point $P_2$ is located at,

$$S^* = \frac{\alpha_{DD}(1 - \Pi) - \alpha_{SD}}{\delta} \qquad (8a)$$

$$D^* = \frac{\alpha_{SS} - (1 - \Pi)\alpha_{DS}}{\delta} \qquad (8b)$$

Here $\delta = \alpha_{DD}\alpha_{SS} - \alpha_{SD}\alpha_{DS}$. This solution is stable when $P > B$, i.e., when

$$\frac{\alpha_{SD}}{\alpha_{DD}} < 1 - \Pi. \qquad (9)$$

If there is no direct influence of one subpopulation on the other, $\alpha_{SD} = \alpha_{DS} = 0$, $S^* = (1 - \Pi)/\alpha_{SS}$, $D^* = 1/\alpha_{DD}$, the population of differentiated cells is formally the same as in $P_1$; symmetric division reduces the population of cancer stem cells. In fact, if $p_d > p_s$, no stable coexistence would be possible, unless cancer stem cells cooperate ($\alpha_{SS} < 0$). On the other hand, if the probability of symmetric division is dominant, $p_d < p_s$, the interaction among stem cells must be inhibitory ($\alpha_{SS} > 0$) to avoid unlimited growth.

Another simple case occurs when each subpopulation can only influence the other subpopulation, $\alpha_{SS} = \alpha_{DD} = 0$. Under these conditions, $S^* = 1/\alpha_{DS}$, $D^* = (1 - \Pi)/\alpha_{SD}$.

Stem cells must inhibit differentiated cell growth ($\alpha_{DS} > 0$), while differentiated cells must inhibit stem cell growth (we consider $p_d < p_s$).

## 3. Fitting the model

A detailed experimental study of tumorsphere growth dynamics was carried out by Chen and coworkers [Chen16], who used three cell lines: SUM159, MCF-7, and T47D. Cultures were started using single cancer stem cells. Average tumorsphere sizes were measured every two days in a 10-day experiment. Steady state populations for these monoclonal tumorspheres contain about 200 cells.

We fitted the results of the experiments described in [Chen16] with our model, using $S(0) = 1$ and $D(0) = 0$ as initial conditions. The results are shown in Fig. 3 and the fitting parameters are presented in Table 1. Note that, despite the large error bars, the model fits very well the data points for the three cell lines (to make a meaningful comparison, we converted spheroid sizes to cell numbers). The first substantial conclusion is that the parameters related to interspecific interactions are always negative, while those related to intraspecific interactions are positive. This implies that cells belonging to different subpopulations cooperate, but cells of the same subpopulations compete. According to Eqs. (7) and (9), under these conditions it is always the coexistence fixed point that is stable. Therefore, the resulting spheroid is expected to be composed of a mixture of stem and differentiated cancer cells.

|  | $\alpha_{SS}$ | $\alpha_{SD}$ | $\alpha_{DS}$ | $\alpha_{DD}$ | $r_S$ |
|---|---|---|---|---|---|
| T47D | 0.0870 | -0.0550 | -0.0150 | 0.0179 | 1.1557 |
| MCF7 | 0.0439 | -0.0287 | -0.0050 | 0.0104 | 1.4478 |
| SUM159 | 0.0463 | -0.0089 | -0.0179 | 0.0054 | 2.5332 |

**Table 1:** Parameter values obtained by fitting model (1) to the data in [Chen16]. The signs of the interaction coefficients $\alpha_{jj}$ indicate that there is interspecific cooperation and intraspecific competition.

The sign of the intraspecific competition is not surprising. On one hand, cells belonging to the same subpopulation compete for the available resources (nutrients, oxygen and space); on the other, CSCs are expected to try to limit, via quorum sensing, their own numbers, as they do in noncancerous tissues. This is likely to be the reason why $\alpha_{SS}$ is in all cases several times larger than $\alpha_{DD}$ and, as a result, the tumorsphere has a larger proportion of differentiated cancer cells. The observed positive influence of the DCCs on the generation of CSCs may be interpreted as evidence of phenotypic plasticity: some DCCs dedifferentiate in regions where CSCs are scarce, contributing to CSC population growth. This would explain the negative sign of $\alpha_{SD}$. CSCs are expected to strengthen their niches. They thus favor the creation of additional differentiated cells, which explains why $\alpha_{DS} < 0$. On the other hand, since the number of stem cells is limited by niche size, stem cell daughters must compete to occupy the niche [Batll17, Morri14]; therefore, we have to include the possibility that some CSCs differentiate into DCCs, a phenomenon that would also contribute to generate negative values of $\alpha_{DS}$.

When there are few experimental points, different parameter combinations may give satisfactory fits [Rodri11]. In our analysis of Chen's data we observed that small variations of the interaction parameters lead to apparently adequate fits to the total population data but to mediocre results for the CSC subpopulation fraction. For this reason, and to confirm the ability of our model to generate reliable predictions, we used an additional result provided by [Chen16]. These authors use a different experiment with the T47D cell line to estimate that the percentage of CSCs at day 14 is 16%, confirming that CSCs and DCCs coexist in the steady state. We include this point as an additional experimental point and represent the results in Fig. 4 (continuous lines). The curves obtained using the parameters in Table I, extrapolated to day 14 (dashed lines), predict a slightly bigger spheroid. The predictions for the DCC and CSC populations are also included. Note that the model overestimates the fraction of CSCs, an error that can be ascribed to the lack of long-time information in the original dataset.

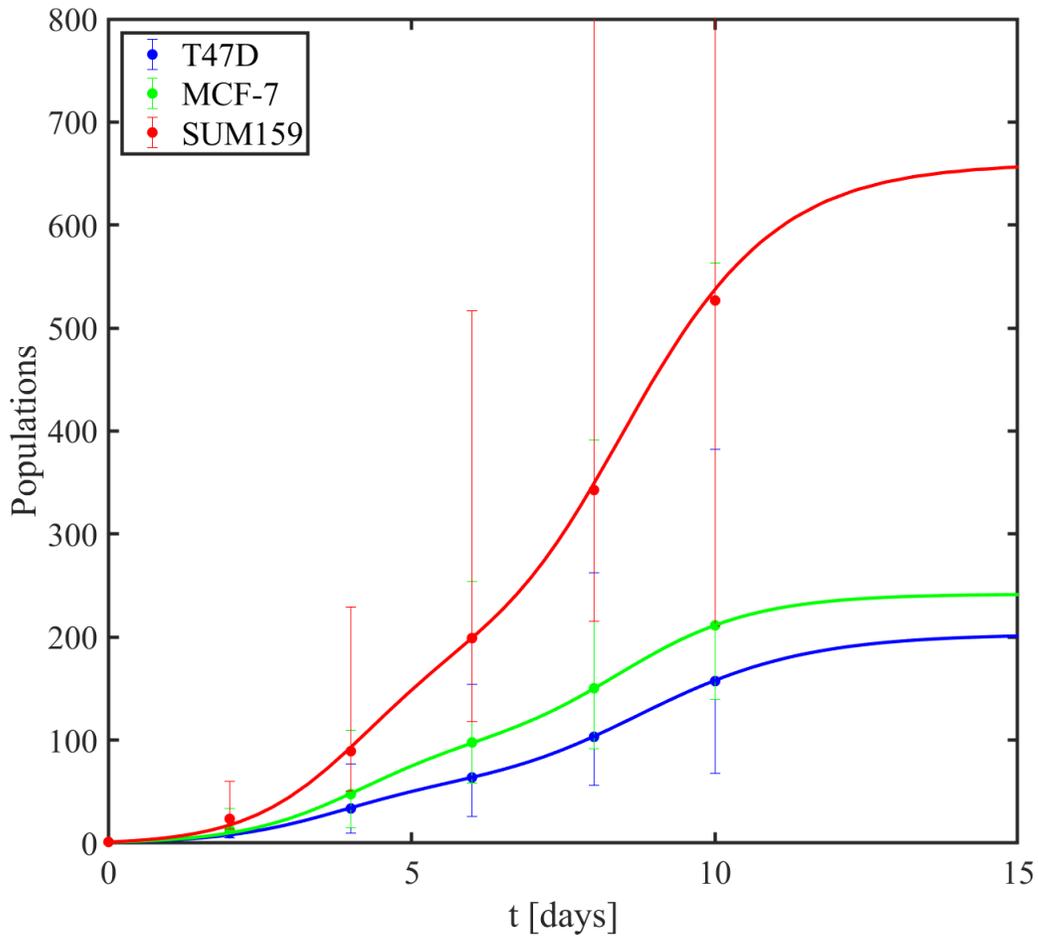

**Fig. 3.** Tumorsphere growth for the indicated cancer cell lines. Datapoints were taken from [Chen16]. Continuous lines are model fits.

| | $\alpha_{SS}$ | $\alpha_{SD}$ | $\alpha_{DS}$ | $\alpha_{DD}$ | $r_S$ |
|---|---|---|---|---|---|
| Predicted | 0.0870 | -0.0550 | -0.0150 | 0.0179 | 1.1557 |
| Obtained | 0.2575 | -0.0897 | -0.0523 | 0.0238 | 1.3256 |
| Relative error | 0.6621 | 0.3868 | 0.7130 | 0.2479 | 0.1282 |

**Table 2**: Comparison between the parameter values obtained by fitting our prediction with 10-day data points ("predicted") and those obtained by including the 14-th day datum ("obtained"). Note the large error bars in Fig. 3. The sizable relative error suggests a large sensitivity to experimental conditions.

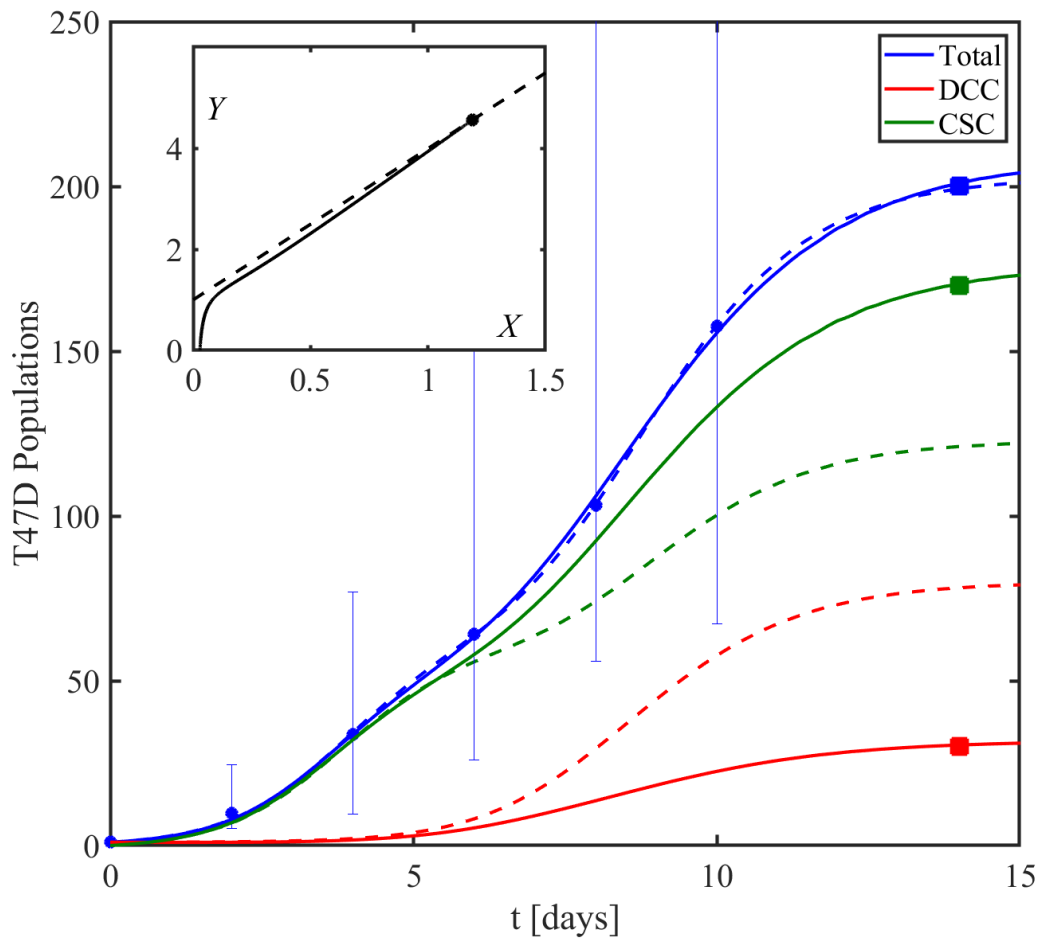

**Fig. 4.** Fits to Chen's results for the T47D line. Continuous lines result from the inclusion of the 14-day point; dashed lines extrapolate the results presented in Fig. 3 and do not include this point. Blue lines correspond to the total spheroid population, green and red lines correspond to the DCC and CSC populations, respectively. There is a single experimental point for the partial populations, which corresponds to the measurements at day 14. Inset, solid line: Trajectory, in phase space, of Chen's experiment [Chen16; dashed

line: Critical point locus. Parameter values obtained by fitting: $P = 0.0663, A = 2.8472,$ and $C = -2.9928$.

## 4. Discussion

The deregulation of normal cellular processes that occurs in cancer has a powerful influence on the evolution of the cell populations. In healthy tissues, differentiated cells secrete factors that inhibit the division and self-renewal of stem cells [Land09, Rodri11]. This inhibition is likely to be perturbed in cancer, where cancer stem cells escape the controls that ensure homeostasis in healthy tissues [Agur10]. In order to gain insights about the reasons for the breakdown of homeostasis that emerges from cancer stem cell activity, we have developed a simple two-population model that describes interspecific and intraspecific interactions between cancer stem cells and differentiated cancer cells. We have shown that this model is suitable to analyze the growth of tumorspheres and may help us to understand the interplay between the cell populations that drive it.

Our analysis of the model indicates that there is a tipping point [Hast18], characterized by a transcritical bifurcation, such that the spheroid resulting from seeding one (or a few) cancer stem cells stops being a homogeneous differentiated cancer cell mass and becomes a two-population mixture. This tipping point occurs when $\alpha_{SD}/\alpha_{DD} = 1 - p_d/p_s$. The right-hand side in this equation contains quantities related to specific cell lines, while the left-hand side is related to cell-cell interactions, which are determined by microenvironmental conditions. Spheroids without stem cells are predicted to be generally smaller than spheroids containing mixed subpopulations. If $\alpha_{SD}/\alpha_{DD} < 1 - p_d/p_s$ and $\alpha_{DD}$ is large enough, the resulting system may be composed of a few $D$ cells or even a single $D$ cell. In this case, an experimenter would conclude that no spheroid has formed. On the other hand, uncontrolled tumorsphere growth has never been observed: growth always stops at a finite size, which also implies that differentiated cancer cells inhibit other differentiated cancer cells (they compete for space, oxygen and nutrients).

It may be argued that similar models may yield equivalent results. To test the model robustness, we have developed a variant such that the stem cell reproduction rate is controlled by all the other cells in the system (see the Appendix). We have not been able to reproduce Chen's results [Chen16] using this modified model, which supports our reliance on the correct design of the model presented here.

From the application of our model to the experiments reported in [Chen16], we conclude that growth is favored by the interspecific interactions: Stem cells promote the increase of differentiated cells that can strengthen their niche, while some differentiated cells may dedifferentiate into stem cells because either they receive perturbed signals from their neighbors or they misinterpret the signals of the microenvironment and phenotypic plasticity runs amok. The possibility of such an altruistic behavior on the part of differentiated cancer cells was noted before [Axel06, Zhou14]. Peer competition explains the inhibitory character of intraspecific interactions.

There is an extensive parameter range for which the number of cancer stem cells temporarily overshoots the equilibrium value, which may or may not contain cancer stem cells. This transient overshooting, which was also observed by Liu and coworkers [Liu13], indicates that cell plasticity plays its most important role in the transient growth dynamics, a finding that agrees with those of [Zhou14]. These authors also remark that reversible phenotype switching between cells increases the collective fitness of the tumor, providing a useful strategy for cancer cell populations defying anticancer mechanisms [Axel06, Zhou14].

Of course, signals in a real tumor are far more complex than those pertaining to *in vitro* systems. But, as remarked by Morrissey and Vermeulen [Morri14, Klein11], "the fundamental dynamics (of stem cell systems) are universal, although the spatial features of the various tissues may be different". We expect that this universality is at least partially conserved in cancer stem cells, which would mean that the basic information obtained from a simple two-population model may help us to interpret the more complex *in vivo* systems. This model could thus be the basis to elaborate an extended version that should include a more complex set of interactions and would let us make more precise predictions about the growth of real cancers and the effect of anti cancer stem cell therapies.

**Acknowledgments**
This work was supported by SECyT-UNC (projects 32720180100257CB and 05/B457) and CONICET (PIP 11220150100644), Argentina We also thank Dr. Luciano Vellón for useful discussions.

**References**

[Agur10] Z. Agur, Y. Kogan, L. Levy, H. Harrison, R. Lamb, O.U. Kirnasovsky, and R.B. Clarke. Disruption of a quorum sensing mechanism triggers tumorigenesis: a simple discrete model corroborated by experiments in mammary cancer stem cells, *Biol Direct* **5**:20, 1-11 (2010).

[Al-Hajj04] M. Al-Hajj and M.F. Clarke. Self-renewal and solid tumor stem cells, *Oncogene* **23**, 7274-7282 (2004).

[Axel06] R. Axelrod, D.E. Axelrod, and K.J. Pienta. Evolution of cooperation among tumor cells, *Proc. Natl. Acad. Sci. U. S. A*. **103**, 13474-13479 [2006].

[Batll17] E. Batlle and H. Clevers, Cancer stem cells revisited, *Nat. Med*. **23**, 1124-1134 (2017).

[Boma07] B.M. Boman, M.S. Wicha, J.Z. Fields, and O.A. Runquist. Symmetric division of cancer stem cells--a key mechanism in tumor growth that should be targeted in future therapeutic approaches, *Clin. Pharmacol. Ther.*, **81**(6), 893–898 (2007).

[Brit03] N. F. Britton, "Essential Mathematical Biology", Springer, London (2003).

[Chen16] Y.C. Chen, P.N. Ingram, S. Fouladdel, S.P. Mcdermott, E. Azizi, M.S. Wicha, and E. Yoon. High-throughput single-cell derived sphere formation for cancer stem-like cell identification and analysis, *Sci. Rep.* **6**:27301, 1-12 (2016).

[Ding08] D. Dingli, A. Traulsen, and J.M. Pacheco. Chronic myeloid leukemia: Origin, development, response to therapy, and relapse, *Clinical Leukemia* **2**, 133-139 (2008).

[Ende09] H. Enderling, A.R. Anderson, M.A. Chaplain, A. Beheshti, L. Hlatky, and P. Hahnfeldt. Paradoxical dependencies of tumor dormancy and progression on basic cell kinetics, *Cancer Research* **69**, 8814–8821 (2009).

[Eramo08] A. Eramo et al., Identification and expansion of the tumorigenic lung cancer stem cell population, *Cell Death Differ.* **15**, 504-514 (2008).

[Ferr12] J.E. Ferrell Jr., Bistability, bifurcations, and Waddington's epigenetic landscape, *Curr. Biol.* **22**, R458 – R466 (2012).

[Fuchs04] E. Fuchs, T. Tumbar, and G. Guasch, Socializing with the neighbors: stem cells and their niche, *Cell* **116**, 769–778 (2004).

[Gang06] R. Ganguly and I.K. Puri. Mathematical model for the cancer stem cell hypothesis, *Cell Prolif.* **39**(1), 3–14 (2006).

[Gao13] X. Gao, J.T. McDonald, L. Hlatky, and H. Enderling. Acute and fractionated irradiation differentially modulate glioma stem cell division kinetics, *Cancer Research* **73**(5), 1481–1490 (2013).

[Ghos17] D. Ghosh, S. Khajanchi, S. Mangiarotti, F. Denis, S.K. Dana, and Ch. Letellier. g How tumor growth can be influenced by delayed interactions between cancer cells and the microenvironment?, *Biosystems* **158**, 17–30 (2017).

[Gu11] Y. Gu, J. Fu, P.K. Lo, S. Wang, Q. Wang, and H. Chen. The effect of B27 supplement on promoting in vitro propagation of Her2/neu-transformed mammary tumorspheres, *J. Biotech Res*. **3**, 7-18 (2011).

[Guis17] N. Guisoni, R. Martínez-Corral, J. García-Ojalvo, and J. de Navascués. Diversity of fate outcomes in cell pairs under lateral inhibition, *Development* **144**, 1177-1186 (2017)

[Guo17] Ch. Guo, S. Ahmed, Ch. Guo, and X. Liu. Stability analysis of mathematical models for nonlinear growth kinetics of breast cancer stem cells, *Math. Methods in the App. Sciences* **40** (14), 5332–5348 (2017).

[Gup11] P.B. Gupta, C.M. Fillmore, G. Jiang, S.D. Shapira, K. Tao, C. Kuperwasser, and E.S. Lander. Stochastic state transitions give rise to phenotypic equilibrium in populations of cancer cells, *Cell* **146**, 633-644 (2011).

[Hajj03] M. Al-Hajj, M.S. Wicha, A. Benito-Hernández, S.J. Morrison and M.F. Clarke. Prospective identification of tumorigenic breast cancer cells, *Proc. Natl. Acad. Sci. USA* **100**, 3983-3988 (2003).

[Hast18] A. Hastings et al. Transient phenomena in ecology, *Science* **361**, eaat6412, 1-9 (2018).

[Klein11] A.M. Klein and B.D. Simons. Universal patterns of stem cell fate in cycling adult tissues, *Development* **138**, 3103 – 3111 (2011).

[Land09] A.D. Lander, K.K. Gokoffski, F.Y.M. Wan, Q. Nie, and A.L. Calof. Cell lineages and the logic of proliferative control, *PLoS Biol.* **7**:e15 (2009).

[Lapi94] T. Lapidot et al. A cell initiating human acute myeloid leukaemia after transplantation into SCID mice, *Nature* **367**, 645-648 (1994).

[Lapor12] C.A.M. La Porta, S. Zapperi, and J.S. Sethna, Senescent cells in growing tumors: population dynamics and cancer stem cells, *PLOS Comp. Biol.* **8**, e1002316 (2012).

[Li07] C. Li, D.G. Heidt, P. Dalerba, C.F. Burant, L. Zhang, V. Adsay, M. Wicha, M.F. Clarke, and D.M. Simeone, Identification of pancreatic cancer stem cells, *Cancer Res.* **67,** 1030-1037 (2007).

[Lin17] M. Lin, A.E. Chang, M. Wicha, Q. Li, and S. Huang. Development and application of cancer stem cell-targeted vaccine in cancer immunotherapy, *J. Vaccines Vaccin.* **8**, 1000371, 1-3 (2017).

[Liu13] X. Liu, S. Johnson, S. Liu, D. Kanojia, W. Yue, U.P. Singh, Q. Wang, Q. Wang, Q. Nie, and H. Chen. Non-linear growth kinetics of breast cancer stem cells: Implications for cancer, *Scientific Reports*, **3**, 2473 (2013).

[Morri14] E.R. Morrissey and L. Vermeulen, Stem cell competition: how speeding mutants beat the rest, *EMBO J.* **33**, 2277- 2278 (2014).

[Obrien07] C.A. O'Brien, A. Pollett, S. Gallinger, and J.E. Dick, A human colon cancer cell capable of initiating tumor growth in immunodeficient mice, *Nature* **445**, 106-110 (2007).

[Rodri11] I.A. Rodríguez-Brenes, N.L. Komarova, and D. Wodarz. Evolutionary dynamics of feedback escape and the development of stem-cell-driven cancers, *Proc. Natl. Acad. Sci.* **108**, 18983-18988 (2011).

[Scad14] D.T. Scadden, Nice neighborhood: Emerging concepts of the stem cell niche, *Cell* **157**, 41-50 (2014).

[Schof78] R. Schofield, The relationship between the spleen colony-forming cell and the haemopoietic stem cell, *Blood Cells* **4**, 7–25 (1978).

[Singh04] S.K Singh et al., Identification of human brain tumour initiating cells, *Nature* **432**, 396-401 (2004).

[Sree18] M. Sreepadmanabh and B.J. Toley. Investigations into the cancer stem cell niche using in-vitro 3-D tumor models and microfluidics, *Biotechnol. Adv.* **36**, 1094-1110 (2018).

[Toma10] C. Tomasetti and D. Levy. Role of symmetric and asymmetric division of      stem cells in developing drug resistance, *Proc. Natl. Acad. Sci. USA,* **107**:16766-71 (2010).

[Viei13] R.V. dos Santos and L.M. da Silva. The noise and the KISS in the cancer stem cell niche, *J. Theor. Biol.* **335**, 79–87 (2013).

[Visv12] J.E. Visvader and G.J. Lindeman. Cancer stem cells: Current status and evolving complexities, *Cell Stem Cell* **6**, 717-728 (2012).

[Wacl15] B. Waclaw, I. Bozic, M.E. Pittman, R.H. Hruban, B. Vogelstein, and M.A. Nowak. Spatial model predicts dispersal and cell turnover cause reduced intra-tumor heterogeneity, *Nature* **525**, 261–267 (2015).

[Weis15] L.B. Weiswald, D. Bellet, and V. Dangles-Marie. Spherical cancer models in tumor biology, *Neoplasia (New York, N.Y.)*, *17*(1), 1–15 (2015).

[Zhou14] D. Zhou, Y. Wang, Y. and B. Wu. A multi-phenotypic cancer model with cell plasticity, *J. Theor. Biol.* **357**, 35-45 (2014).

## Appendix

Here we develop a variant of the model of Eq. (1), where we consider that the effective number of stem cell divisions is controlled not only by the intrinsic cell dynamics but also by the influence of the tumor tissue as a whole.

$$\frac{dS}{dt} = r_s(p_s - p_d)[1 - (\alpha_{SS}\ S + \alpha_{SD}\ D)]S \qquad (A1a)$$

$$\frac{dD}{dt} = \{r_D D + r_S S[1 - (p_s - p_d)]\}[1 - (\alpha_{DD}D + \alpha_{DS}S)] \qquad (A1b)$$

Non-dimensionalization: The definitions of X, Y, $\tau$, and P are the same as in Eq. (2) in the main text. Now we redefine the parameters A, B, and C,

$$A = \frac{\alpha_{SS} r_D P}{\alpha_{DD} r_s (1 - p)}$$

$$B = \frac{\alpha_{SD} r_D P}{\alpha_{DD} r_S (1-p)} \qquad (A2)$$

$$C = \frac{\alpha_{DS} r_D}{\alpha_{DD} r_S (1 - p)}$$

These transformations lead again to Eqs. (3); consequently, we get the same set of fixed points.

- The trivial point $P_0 = (0, 0)$ is unstable when P > 0.
- The DCC fixed point at $P_1 = (0, 1)$, which is stable as long as P < B, which is equivalent to $\alpha_{DD} < \alpha_{SD}$.
- The coexistence fixed point, at the location defined by Eq. (4)

Even if the location of the fixed points coincides with those of the model described in the paper, the biological implications are different. Here the bifurcation occurs when $\alpha_{DD} = \alpha_{SD}$. It does not depend on $p_d$ or $p_s$. All our attempts to fit Chen's data [Chen16] using this model failed.